\documentclass[a4paper,10pt]{article}
\usepackage{amsmath}
\usepackage{amssymb}
\usepackage{a4wide}
\usepackage[english]{babel}
\usepackage[T2A]{fontenc}
\usepackage{amsmath}
\usepackage[cp1251]{inputenc}
\usepackage{graphicx}

\begin{document}

\centerline{\Large \bf Chiral Vortical Effect in Superfluid}

\vspace{10mm}

\centerline{V.P. Kirilin, A.V. Sadofyev, V.I. Zakharov}

\vspace{5mm}

\centerline{Institute for Theoretical and Experimental Physics, Moscow}
\date{}


\begin{center}{\bf Abstract}\end{center}
We consider rotating superfluid pionic liquid,
with superfluidity being induced by isospin chemical potential.
The rotation is known to result in a chiral current flowing along 
the axis of the rotation. We argue that in case of superfluidity
the chiral  current  is realized on fermionic zero modes propagating
along vortices. The current evaluated in this way differs by a factor 
of two  from the
standard one. The reason is that the chiral charge is carried by
zero modes which propagate with speed of light,
and thus the liquid cannot be described by a single (local) velocity, like it is assumed in standard derivations.



\section{Introduction}
Recently, there were intense studies of hydrodynamics of
chiral liquids. A crucial  novel point  is  existence of  
new transport  coefficients, overlooked in the text-book approaches.
In particular, it was discovered \cite{erdmenger,sonsurowka}
that in case of liquids with chiral constituents
there exists chiral current $j^5_{\mu}$ proportional  to  the  vorticity $\omega^{\mu}$:
\begin{eqnarray}\label{vortical}
\delta j^5_\mu = \frac{\mu^2}{2\pi^2}\omega_\mu~,
\end{eqnarray}
where $\omega^\mu = \frac{1}{2} \epsilon^{\mu\nu\alpha\beta} u_{\nu}\partial_\alpha u_\beta $, $\mu$ is the chemical potential and $u_{\mu}$ is the 4-velocity of an element of the liquid. The observation (\ref{vortical}) can be considered as a kind 
of generalization of the chiral magnetic
effect (CME) known since longer time, 
for review and further references see, e.g., \cite{kharzeev}. In the latter
case the chiral current is directed along the magnetic field in the local rest frame:
\begin{equation}\label{magnetic}
\delta j^5_\mu~=~\frac{\mu}{2\pi^2}q B_\mu~,
\end{equation}
where
$B^\mu \equiv  \epsilon^{\mu\nu\alpha\beta} u_{\nu}\partial_\alpha A_\beta$,
$q$ is the charge of the constituent fermion.

The currents (\ref{vortical}), (\ref{magnetic}) represent macroscopic manifestations
of the triangle anomaly in  the  underlying chiral theory. 
It is a fascinating observation that quantum, loop effects 
manifest in the hydrodynamic, that is  classical approximation. 
Naturally enough, Eqs (\ref{magnetic}), (\ref{vortical}) were considered within
various frameworks.
Originally,  Eq. (\ref{vortical})
was derived by considering the entropy current \cite{sonsurowka}.
Later, it was obtained in other ways, 
 in particular, in the language of anomalies in
effective theory in the presence of a chemical potential and using other approaches \cite{sonzhitnitsky,zahed,sadofyev,zakharov,Tigran,Roy}. Most recently \cite{new} the assumptions
needed to derive (\ref{vortical}), (\ref{magnetic}) were reduced to general Ward identities of  the
underlying microscopic theory. 
 
 Despite of variety of the assumptions tried, all these approaches treat
the liquid as a slowly-varying in its properties
medium.  Since the constituents of the liquid are chiral
this assumption might look, though, not well justified. Indeed, the occupation numbers
for fermions are never large and it is not a priori clear
how to introduce hydrodynamics, or classical approximation for fermions.
  This problem can be addressed within
  another approach which starts with 
 a microscopical picture
and the central role  is played  then by low-dimensional defects.
 This approach
goes back to papers   \cite{goldstonewilczec,nielsen,callanharvey}
where
it was demonstrated that defects in field theory are   closely tied to the realization of anomaly. In particular, anomaly in 2n+2 dimensional theory is connected with 2n dimensional index density and can be understood in terms of fermionic zero modes on strings 
and domain walls  \cite{callanharvey}.
In all the cases the chiral current is carried by fermionic zero modes
living on the defects.

 One can expect, therefore, that microscopically the anomaly
 is realized on  vortex-like strings while 
the continuum-medium results (\ref{vortical}), (\ref{magnetic})
 arise  upon averaging  over a large number of defects.
In case of the chiral magnetic effect (see Eq (\ref{magnetic}))
such a mechanism was considered, in particular,  in Refs. \cite{zhitnitsky}, \cite{kharzeev}
 and
the final result (\ref{magnetic}) is reproduced on the microscopic level
as well. 

In Refs. \cite{zhitnitsky}, \cite{kharzeev} the vortices are modelled by regions of space free of the medium substance.
In case of superfluidity the vortices are much better  understood
dynamically.
The crucial point is that
 the velocity field of the superfluid is known to be potential, and, naively, the vorticity vanishes everywhere. If this were true, the chiral current
 (\ref{vortical}) would vanish.  But it is well known, of  course, that the angular
momentum is still transferred to the liquid through vortices.
The potential is singular on the linear defects, or quantum vortices.
The vorticity 
is not vanishing
on these defects and the entire  chiral current flows through 
the vortices. Although the vorticity locally vanishes everywhere but the defects,
globally, the whole  liquid can be regarded as rotating, e.g. it possesses
angular momentum (provided that the angular velocity  is large enough) \cite{landau6}.
 This is an example of recovering the continuum limit upon averaging
 over a large number of defects.
 
In the microscopic picture, the evaluation of chiral effects reduces to counting
zero chiral modes. The calculation is in two steps in fact.
In case of the chiral magnetic effect the technique was elaborated  
in Ref. \cite{nielsen} where details can be found. 
First, one considers plane pierced by magnetic field $B$. Then there is an index theorem which relates the number of zero modes to the magnetic flux:
\begin{equation}
N_{0,\perp}~\sim ~q\int d^2x B~,
\end{equation} 
where $q$ is the charge of the particle.
These zero modes exist for any value of the longitudinal momentum directed along the
magnetic field. At the second step one integrates over
the longitudinal momentum up to the chemical potential $\mu$.
Combining all the factors one gets for the total number of zero modes
relevant to the chiral magnetic effect:
\begin{equation}
N_0~\sim q\cdot B\cdot \mu
\end{equation}

Proceed now
to the vortical chiral effect (CVE). The evaluation of the number of zero modes is the same
two-step process as above. However, there is no external electromagnetic field $A_{\mu}$
any longer. Instead, one considers 4-velocity of the liquid $u_{\mu}$ as an external field.
One obtains the number of zero modes $N_{0,\perp}$ through the substitution:
\begin{equation}
qA_{\mu}~\to~\mu u_{\mu}.
\end{equation}   
In this sense  the chemical potential $\mu$ plays the role of the strength of
interaction (like the charge $q$).
The next step is essentially unchanged. One integrates over the 
longitudinal  momentum 
up to the chemical potential, $p_{\parallel}\le \mu$ 
As a result, the total number of zero modes relevant to the
chiral vortical effect is  proportional to $\mu^2$. 
One factor of $\mu$ comes from the flux of the vorticity (analogy to the
flux of the magnetic field). The origin of the other factor $\mu$ is
the integration over the longitudinal momentum.

The result for the chiral currents obtained through counting
zero modes can be compared to the  evaluation of the same currents within effective
field theory. In both cases it is the triangle graphs
which control the effect. The chiral magnetic effect is linear both in the interaction 
$qA_{\mu}$ and $\mu u_{\mu}$. The chiral vortical effect is
quadratic in $\mu u_{\mu}$. In both cases the triangle graphs
reproduce the resuts (\ref{magnetic}). (\ref{vortical}).

 Our central point is that 
 the chiral current evaluated in terms of the zero modes
differs from (\ref{vortical}) by a factor
of two. The difference can be traced back to the fact that the fermionic
zero modes propagate with speed of light and are not equilibrated to
the local 4-velocity $u_{\mu}$ of an element of the  liquid.
If we consider magnetic field in charged superfluid then we  find that
the magnetic field plays the role similar to the vorticity and there are  strings carrying magnetic flux, and, therefore, zero modes. The chiral magnetic effect in terms of  the zero modes turns to be the same  Eq. (\ref{magnetic}), echoing 
the results of \cite{zhitnitsky}, \cite{kharzeev}.
In this sense, the vortical and magnetic chiral effects in  superflluid 
substantially differ from each other. We will further comment on this
difference in the conclusions.

  The outline of the paper is as follows: we will start by recalling some common points regarding superfluidity, and the way it arises in quark medium via chiral Lagrangian. We also remind the reader the derivation of (\ref{vortical}), (\ref{magnetic}). In the next section, we introduce the interaction between fermions and the Goldstone field, and derive the chiral current via anomaly. We then proceed to the microscopic derivation of the current in terms of zero modes. In the last section we discuss our results and mention a few open questions.

\section{Hydrodynamics, vorticity and superfluid}
We start our detailed considerations by recalling how   superfluidity 
arises in pion medium at non-zero isospin chemical potential and zero temperature
\cite{sonstephanov}. The pion medium 
at zero temperature is described by chiral Lagrangian  and at non-zero isospin chemical potential $\mu_I$ it takes the form:

\begin{eqnarray}
\label{lag}
L=\frac{1}{4}f_\pi^2Tr[D^{\mu}U(D_{\mu}U)^\dag],
\end{eqnarray}
where $D_0U=\partial_0 U-\frac{\mu_I}{2}[\tau_3, U]$, $D_iU=\partial_iU$. Here for simplicity we consider zero quark masses. The chiral symmetry is spontaneously broken to $SU(2)_{L+R}$. 
Moreover, at non-zero $\mu_I$ this symmetry is explicitly
broken to $U(1)_{L+R}$. And, finally, the $U(1)_{L+R}$ symmetry is spintaneously
broken, triggering supefluidity. The argumentation is as follows, The potential energy in (\ref{lag}) equals to

\begin{eqnarray}
V_{eff}(U)=\frac{f_\pi^2\mu_I^2}{8}Tr[\tau_3U\tau_3U^\dag-1],
\end{eqnarray}
minima of that potential can be captured by substitution $U=\cos{\alpha}+i(\tau_1\cos{\phi}+\tau_2\sin{\phi})\sin{\alpha}$:

\begin{eqnarray}
V_{eff}(\alpha)=\frac{f_\pi^2\mu_I^2}{4}(\cos{2\alpha}-1)
\end{eqnarray}
and for the minimum one readily obtains $\cos{\alpha}=0$. Then, depending on sign of $\mu_I$, squared mass of either $\pi^+$ or $\pi^-$ becomes negative,
signalling the condensation of the corresponding pion field. This means that the vacuum is described by $U = i(\tau_1\cos{\phi}+\tau_2\sin{\phi})$ instead of the "usual" vacuum $U = I$, and the emergence of the new order parameter $\langle \overline{u} \gamma_5 d\rangle + h.c. = 2 \langle\overline{\psi}\psi\rangle_{vac} \sin(\alpha) = 2 \langle \overline{\psi}\psi\rangle_{vac}$. The system is thus a charged superfluid.  Degeneracy with respect to the $\phi$ angle indicates that $\phi$ can be identified with Goldstone boson, and that the symmetry $U(1)_{L+R}$ is spontaneously broken. More accurately, one needs to make a replacement, $\phi + \mu t \rightarrow  \phi$, since $\phi$ enters stress-energy tensor as a combination $\partial_0 \phi + \mu$;  here and thereafter we consider $\phi$ as redefined. In addition to the massless Goldstone mode, there are two massive modes. Note that these results can be reproduced holographically \cite{zamaklar}.

 To study vortices  we rely on the hydrodynamic approximation. The hydrodynamics of
 a charged superfluid incorporating the Goldstone field $\phi$
 is worked out in
 \cite{sonsup} and references therein. The symmetry associated with
   the charge  is
 spontaneously broken by a condensate
 and the Goldstone field $\phi$ is added to the standard hydrodynamical variables. The ground state is described by the $ \partial_0 \phi = \mu$ (Josephson equation), which implies non-vanishing charge density. For a general motion of the superfluid the corresponding velocity is given by $u_\mu^s = \partial_\mu \phi / |\partial_\mu \phi|$, $v^s_i = \partial_i \phi/\mu$ (non-relativistically)  and one readily finds that  $\mathbf{rot}~ \mathbf{v}^{s}=0$. This seems to forbid rotational motion.

 However, it is known that for superfluid (at $T = 0$, so that normal component is absent) put into a rotating bucket the solution with non-zero angular momentum is energetically preferable for angular velocities larger than some critical value $\Omega > \Omega_c$. This implies non-zero velocity field circulation, which means that the velocity potential, given by the Goldstone field, is multi-valued. The Goldstone field is then ill-defined on a linear defect called the vortex line, and is given by $\phi = \mu t + n\cdot \varphi$ in
 the limit of vanishing thickness of the core \cite{Landau9}. Note that this description is valid for distances $r >> a$ where $a$ is the size of the vortex. The value of critical angular velocity is given by $\Omega_c = \frac{1}{\mu R^2} \log\frac{R}{a}$ where $R$ is the radius of the cylindrical bucket \cite{Landau9}. For higher angular velocities the  circulation is increased via the generation of additional vortices with $n = 1$, which are energetically favorable to $n > 1$ vortices. This, and the fact that due to mutual repulsion, vortices tend to be distributed uniformly, causes the motion of the liquid induced by the vortices to imitate uniform rotation, $ \mathbf{v} = [\mathbf{\Omega}\mathbf{r}]$ for high enough angular velocities $\Omega >> \Omega_c$. This observation allows to compare the chiral vortical effect evaluated microscopically with
 (\ref{vortical}) obtained macroscopically. 

In next section we consider chiral vortical effect and here it is worth mentioning how it is obtained. We follow \cite{zakharov} and start with action for fermion at non-zero chemical potential(here for simplicity we consider the case of one flavor):

\begin{equation}
S = \int d^4x \left(i\overline{\psi} \gamma^\mu(\partial_\mu-iqA_\mu)\psi + \mu u_\mu\overline{\psi}\gamma^\mu\psi\right),
\end{equation}
where the substitution $\mu\gamma^0\rightarrow\mu u_\mu\gamma^\mu$ makes notations relativistically invariant.  One can obtain CVE and CME as anomaly induced currents. At zero axial chemical potential one should expect chiral effects only in the axial current. After calculating of anomaly we get:

\begin{eqnarray}
\partial_\mu \overline{\psi}\gamma^\mu\gamma_5\psi=-\frac{1}{4\pi^2}\epsilon_{\mu\nu\alpha\beta}\left(\partial^\mu
(A^\nu+\mu u^\nu)\partial^\alpha (A^\beta+\mu u^\beta)\right).
\end{eqnarray}
 Alternatively, one can introduce a modified  current:

\begin{eqnarray}
j^5_\mu=\overline{\psi}\gamma_\mu\gamma_5\psi+\frac{1}{2\pi^2}\mu^2\omega_\mu+\frac{1}{2\pi^2}\mu q B_\mu,
\end{eqnarray}
which satifies non-anomalous (in case $\mathbf {E \cdot B} = 0$) equations 
and in hydrodynamical limit one  replaces $\overline{\psi}\gamma_\mu\gamma_5\psi\rightarrow n_5u_\mu$. Note the
factor of two difference  between the coefficients in front of
the chiral-vortical and chiral-magnetic effects.
This is a  consequence of identity of two vertices in the anomaly triangle diagram in the CVE case. For CVE case that identity takes place and one should divide the result of diagram on two - $\frac{\mu^2}{4\pi^2}\epsilon_{\mu\nu\alpha\beta}u^\nu\partial^\alpha u^\beta$ in contrast to the case of CME - $\frac{\mu}{2\pi^2}\epsilon_{\mu\nu\alpha\beta}u^\nu\partial^\alpha A^\beta$.

\section{Chiral Currents Via Anomaly}
We consider quark matter at finite isospin chemical potential, which forms a  superfluid  (see \cite{sonstephanov})  with the corresponding NG boson $\phi$. It is argued in \cite{sonsup} that $\partial_\mu\phi$ can be identified with non-normalised superfluid velocity. The vortex configuration is in principle determined by the angular velocity. We address a general situation, when the $n$, the quantum number of circulation, is rather high (but not high enough to  ruin the superfluidity). We mentioned above that an energetically preferable configuration is the uniform distribution of vortices with $n = 1$. Nearby any given vortex the Goldstone field is given by \cite{nicolis}:

\begin{eqnarray} \label{vortex}
\phi=\mu t+\varphi,
\end{eqnarray}
where $\varphi$ is the polar angle in the plane orthogonal to the vortex. We will assume that vortices are far one from another $\delta x >> a$, we will calculate the current for a single vortex and then sum it over all vortices, that is simply multiply by $n$.

It is then tempting to introduce the following Lagrangian for the interaction of fermions with the NG boson (we will limit ourselves  to the case of single fermion)
\begin{eqnarray}
L=\overline{\psi}i(\partial_\mu+i \partial_\mu\phi)\gamma^\mu\psi.
\end{eqnarray}
We remind here the following non-relativistic substitution $\partial_0\phi \rightarrow \mu, \partial_{i}\phi \rightarrow \mu v^{s}_{i}$, which relates $\phi$ to the non-normalized potential for the superfluid velocity.
Using standard methods of evaluating the anomalous triangle diagrams one obtains for the axial current (see \cite{goldstonewilczec}, \cite{sonzhitnitsky}, \cite{zahed})

\begin{eqnarray}
j^5_\mu=\frac{1}{4\pi^2}\epsilon_{\mu\nu\alpha\beta}\partial^{\nu}\phi \partial^{\alpha}\partial^{\beta}\phi ,
\end{eqnarray}
This current seems to vanish identically. However,  for the vortex field $\phi = \mu t + \varphi$, and hence:

\begin{eqnarray}
j^5_3 = \frac{\mu}{2\pi}\delta(x,y)
\end{eqnarray}
since $[\partial_x,\partial_y]\phi=2\pi \delta(x,y)$ the total current 
(the sum over the vortices) equals to

\begin{eqnarray}
\label{inanomaly}
J^5_3 = \int d^2x j^5_3 = \frac{\mu}{2\pi}n.
\end{eqnarray}

There is an apparent contradiction between (\ref{inanomaly}) with linear dependence on the chemical potential and (\ref{vortical}) where it is quadratic. It is resolved by noting that $n$, the quantum number for vorticity, depends on $\mu$ \footnote{See \cite{kharzeev}, where $\frac{e \Phi}{2 \pi}$ is simply flux measured in units of flux quantum}. In order to see this, we have to average over the defects to obtain continuum limit as we mentioned above.  Then $n$ has the following continuum limit form:
\begin{eqnarray}
n = \frac{1}{2\pi}\int d x_{i}~ \partial_{i} \phi = \frac{\mu}{2\pi}\int d x_{i} v^{s}_{i} =  \frac{\mu}{\pi}\int d^2 x \omega^3,
\end{eqnarray}
since we can replace $\mathbf{rot}~ \mathbf{v}^{s}= 2 \mathbf{\omega}$. Thus, from (\ref{inanomaly}) one can prescribe an average current density exactly matching (\ref{vortical}).  

This result for CVE conforms with usual result for CME in sense that if we considered charged superfluid and turned on magnetic field, a defect would appear, the current would be obtained by substituting the vortex configuration in the usual formula for CME, and would be concentrated on the vortex. However, we show in following section  that answers obtained through zero modes calculation are different for CVE and CME.

\section{Zero modes}
 We now proceed  to  the microscopic picture based on the zero modes. Our 
 considerations in this section are  close to those of Ref. \cite{zhitnitsky} but there are several significant differences as well. We can write Lagrangian for our system with coupling to
 the Goldstone field $\phi$ in the  following form:

\begin{eqnarray}
L=\overline{\psi}i(\partial_\mu+i \partial_\mu\phi)\gamma^\mu\psi
\end{eqnarray}
Such coupling with the Goldstone field corresponds in non-superfluid limit to naive relativisation $\overline{\psi}\gamma^\mu\psi\partial_\mu\phi\rightarrow \mu u^\mu \overline{\psi}\gamma^\mu\psi$.

 Nearby any given vortex the Goldstone field is given by (\ref{vortex}).
 More important, irrelevant of the details of the configuration, the integral $\int dx_{i} ~ \partial_{i}\phi = 2\pi n$.

Further calculation is close to the one performed in \cite{zhitnitsky}, with a substitution    $A_{i} \rightarrow \partial_i \phi$.
The Hamiltonian has the form 
\begin{eqnarray}
H = -i (\partial_i - i  \partial_i \phi)\gamma^0 \gamma^i 
\end{eqnarray}

then the Dirac equation decomposes:
\begin{eqnarray}
-H_R \psi_L = E \psi_L\nonumber\\
H_R \psi_R = E \psi_R,
\end{eqnarray}
here $H_R = (-i\partial_{i} + \partial_{i} \phi) \sigma_{i}$.
Hence, any solution $\psi_R$ of $H_R \psi_R = \epsilon \psi_R$ simultaneously generates a solution with $E = \epsilon$,
$
\psi=\left(\begin{array}c
0\\
\psi_R
\end{array}\right)
$ and a solution with $E = -\epsilon$,
$
\psi=\left(\begin{array}c
\psi_{R}\\
0
\end{array}\right)
$.

Due to invariance with the respect to translations in the $z$ direction, we decompose $\psi$ using the momentum eigenstates $-i \partial_3 \psi_R = p_3\psi_R$ (it is convenient to take the z-direction periodic with length $L$, and take limit $L \rightarrow \infty$ at the end of the calculation. For each $p_3$ we can write
\begin{eqnarray}
H_R &=& p_3 \sigma^3 + H_\perp \nonumber\\
H_\perp &=& (-i \partial_a - \partial_a \phi) \sigma^a, \;\; a = 1,2.
\end{eqnarray}
Notice that $\{\sigma^3, H_\perp\} = 0$. Hence, if $|\lambda\rangle$ is a properly normalised eigenstate of $H_\perp$ with eigenvalue $\lambda$, then $\sigma_3 |\lambda\rangle$ is a properly normalised eigenstate with eigenvalue $-\lambda$. This means that all eigenstates of $H_\perp$ with non-zero eigenvalue are of the form $|\lambda\rangle$, $|-\lambda\rangle = \sigma_3 |\lambda\rangle$, with $\lambda > 0$. Also, $\sigma_3$ maps zero eigenstates of $H_\perp$, so all eigenstates of $H_\perp$ can be classified with respect to $\sigma_3$.

 We can now express eigenstates of $H_R$ in terms of eigenstates of $H_\perp$. Since $[H_R,{H_\perp}^2] = 0$, $H_R$ will only mix states $|\lambda\rangle$, $|-\lambda\rangle$. For $\lambda > 0$, one can write,
 \begin{eqnarray}
 \psi_R = c_1|\lambda\rangle + c_2 \sigma_3|\lambda\rangle,
 \end{eqnarray}
 where $c_1, c_2$ satisfy:
 \begin{eqnarray}
 \left(\begin{array}{cc} \lambda & p_3 \\
 p_3 & -\lambda \end{array}\right) \left(\begin{array}{c} c_1 \\
 c_2\end{array}\right) = \epsilon \left(\begin{array}{c} c_1 \\
 c_2\end{array}\right).
 \end{eqnarray}
 Thus, $\epsilon = \pm \sqrt{\lambda^2 + p_3^2}$ and,
 \begin{eqnarray}
  \label{cs} \left(\begin{array}{c} c_1\\
 c_2\end{array}\right)_{\pm} = (4(\lambda^2 +
 p_3^2))^{-\frac{1}{4}} \left(\begin{array}{c} \pm
 sgn(p_3) ((\lambda^2 + p_3^2)^{\frac{1}{2}} \pm \lambda)^{\frac{1}{2}} \\
 ((\lambda^2 + p_3^2)^{\frac{1}{2}} \mp
 \lambda)^{\frac{1}{2}}\end{array}\right).
 \end{eqnarray}
 This means that every eigenstate of $H_\perp$ with eigenvalue $\lambda > 0$ produces two eigenstates of $H_R$, while the zero modes of $H_\perp$ are
 eigenstates of $H_R$ with eigenvalue:
 \begin{eqnarray}
 \epsilon = p_3\sigma^3.
 \end{eqnarray}

 Thus, the zero modes of $H_\perp$ are gapless modes of $H$, capable of travelling  up or down the vortex, depending on the sign of $\sigma_3$ and
 chirality. These will be precisely the carriers of the axial current along the vortex.  Let $N_+$ and $N_-$ be the numbers of zero modes with $\sigma_3 = 1$ and $\sigma_3 = 1$,
 respectively. Consider the zero mode of $H_\perp$, $|\lambda\rangle = (u,~ v)$, where $u$ and $v$ are c-functions satisfying
 \begin{eqnarray}
 {\cal D} v = 0, ~~~ {\cal D}^{\dag} u = 0,
 \end{eqnarray}
where

 \begin{eqnarray}
 {\cal D} = -i \partial_1 - \partial_2  -  (\partial_1 \phi - i \partial_2 \phi).
 \end{eqnarray}
 Hence $N_+ = dim(ker({\cal D}^{\dag}))$, $N_- = dim(ker({\cal D}))$ and 

 \begin{eqnarray}
  N = Index(H_\perp) = N_+ - N_- =  dim(ker({\cal D}^{\dagger})) - dim(ker({\cal D})).
 \end{eqnarray}

Note that $H_\perp$ is an elliptic operator. Its index has been computed via various approaches in papers \cite{Casher,semenoff}. In our case the index is given by
\begin{eqnarray}
N = \frac{1}{2\pi}\int d x_{i} ~\partial_i \phi = n.
\end{eqnarray}
(For $n = 1$ the zero mode is easy to construct, see Appendix.)

One can observe that this result is obtained by direct substitution of $e A_{i} \rightarrow \partial_{i} \phi$ in the well known case of magnetic field
parallel to z-axis and uniform in that direction, where the index is given by \cite{zhitnitsky,Casher}
\begin{eqnarray}
N = \frac{e}{2\pi}\int d x_{i} A_{i} = \frac{e}{2\pi} \int d^2 x  B^3.
\end{eqnarray}
Notice, that for the case of superfluid, which we discuss here, the index is essentially an integer.

We now proceed to the computation of the fermion axial current at a
 finite chemical potential $\mu$. The axial current density in the third direction
is given by:
\begin{eqnarray}
j^3_5(x) = \overline{\psi}(x)\gamma^3 \gamma^5 \psi(x) = \psi_L^{\dag} \sigma^3 \psi_L(x) +\psi_R^{\dag} \sigma^3 \psi_R(x).
\end{eqnarray}
We are interested in the expectation value of the
 axial current along the vortex, $J^3_5 =\int d^2x \langle j^3_5(x)\rangle$ and at finite chemical potential, we have:
\begin{eqnarray}
\langle j^3_5(x)\rangle = \sum_{E} \theta (\mu - E)\,\psi^{\dagger}_E(x) \gamma^0 \gamma^3 \gamma^5 \psi_E(x)= \sum_{\epsilon} (\theta(\mu - \epsilon) + \theta(\mu + \epsilon)))\psi_{R\epsilon}^{\dagger}(x)\sigma^3 \psi_{R\epsilon}(x),
\end{eqnarray}
here, $\theta (\mu - E)$ is the Fermi-Dirac distribution (at zero temperature), $\psi_E$ are eigenstates of $H$ with eigenvalue $E$, $\psi_{R \epsilon}$ are eigenstates of $H_R$ with eigenvalue $\epsilon$. By substitution of the explicit form of $\psi_{R\epsilon}$ in terms of $H_\perp$ eigenstates, one obtains:
\begin{eqnarray}
\langle J^3_5\rangle &=&
\frac{1}{L}\sum_{p_3}\sum_{\lambda
>0}\sum_{s = \pm}(\theta(\mu - (\lambda^2 + p_3^2)^{\frac{1}{2}}) +\theta(\mu + (\lambda^2 + p_3^2)^{\frac{1}{2}}))\langle\psi^s_R(\lambda,p_3)|\sigma^3|\psi^s_R(\lambda,p_3)\rangle+\notag\\ &+& \frac{1}{L}\sum_{p_3}\sum_{\lambda = 0} (\theta(\mu - p_3)+\theta(\mu + p_3)) \langle
\lambda |\sigma^3|\lambda\rangle.
\end{eqnarray}
 Here $\lambda > 0$ enumerate eigenstates of $H_\perp$, which generate eigenstates of $H_R$, $\psi^\pm_R(\lambda,p_3)$ with momentum $p_3$
and eigenvalue $\epsilon_{\pm} = \pm \sqrt{\lambda^2 + p_3^2}$, and $\lambda = 0$ label the zero modes of $H_\perp$. An explicit calculation gives
$\langle\psi^s_R(\lambda,p_3)|\sigma^3|\psi^s_R(\lambda,p_3)\rangle = s p_3(\lambda^2 + p_3^2)^{-\frac{1}{2}}$. Consequently, since the result is odd in both $s$ and $p_3$, the sum over all non-zero eigenstates vanishes, and only zero modes of $H_\perp$ generate $J^3_5$. For the zero modes, $\langle \lambda|\sigma^3|\lambda\rangle = \sigma^3$, so we obtain:

\begin{eqnarray}\label{upperlimit}
\label{inmode}
J^3_5 =
(N_+ - N_-) \frac{1}{L}\sum_{p_3}(\theta(\mu - p_3) +
\theta(\mu + p_3))
= n  \int \frac{dp_3}{2\pi} (\theta(\mu - p_3) + \theta(\mu+ p_3)) = \frac{\mu}{\pi} n.
\end{eqnarray}
This result is similar in structure to  (\ref{inanomaly}) but differs from it by a factor of two. The origin of this difference is discussed in conclusions.

The macroscopic calculation through anomaly and the microscopic one through zero modes seem to be of rather different nature. The triangle diagram for the anomaly requires a particular regularization scheme whereas integrals over momenta of zero modes are cut by Fermi - Dirac distribution. It is worth emphasizing that the two answers coincide with each other for the chiral magnetic effect . It means that the observed discrepancy for vortical current indicates the difference between physical approaches rather than the regularization details. That fact justifies comparing of macroscopic and microscopic results for CVE.  

\section{Discussion. Conclusions}
In this note we considered the chiral vortical effect in superfluid.
Considering supefluidity helps to fix the dynamics to a great extent.
The central point is that vorticity is non-vanishing only on linear defects
and, therefore, introducing defects is a kind of a must. Then the picture with defects is only one consistent. Moreover,
the vortices are quantized and this fixes the index which determines 
the number of zero modes, which, in turn, are responsible for the 
chiral current. The quantization of the vortices also accounts for the apparent discrepancy between the usual form of vortical effect (\ref{vortical}) and the form (\ref{inanomaly}), since the vorticity quantum number $n$ depends on $\mu$ (in the same way as the flux quantum in superconductivity depends on the charge e and magnetic field). On the other hand, it is known that upon averaging 
over the defects the rotating superfluid looks the same as ordinary liquid
and one can compare the result for the current with the continuum limit
(\ref{vortical}).

\begin{figure}[ht]
  \includegraphics[width=15cm]{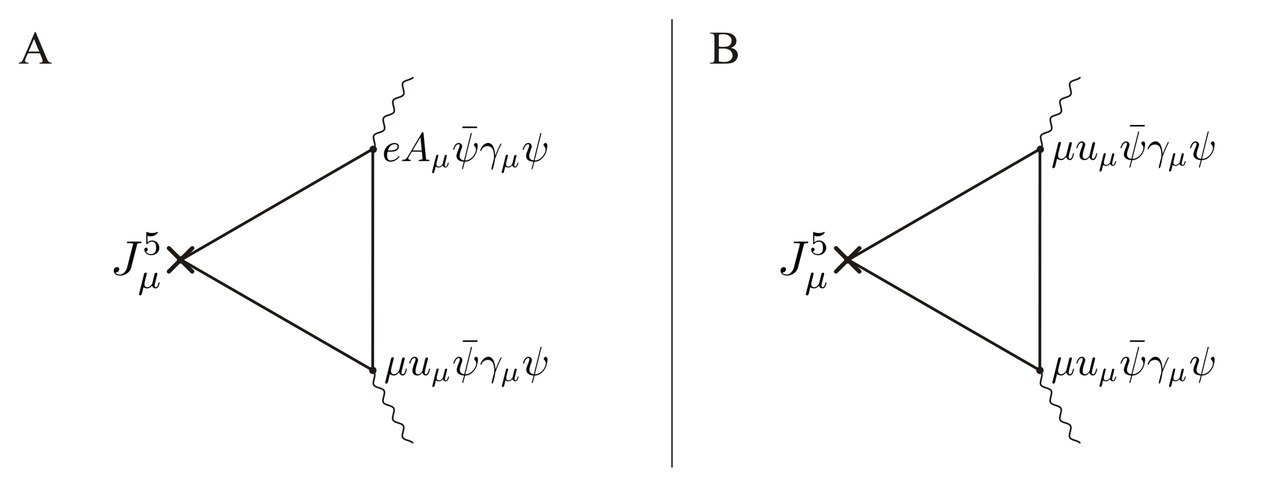}
  \caption{A - contribution to CME, B - contribution to CVE}
   \label{fig1}
\end{figure}

A crucial point is that microscopically we get a factor of two larger value
of the current. To elucidate the origin of this factor it is useful to
compare the triangle anomalous graphs for the chiral-magnetic and
chiral-vortical effects, see Fig\ref{fig1}. The vertices in the graphs are determined
by the corresponding terms  in the (effective) Lagrangian:
\begin{equation}\label{effective}
L_{int}~=~\mu u_{\mu}\bar{\psi}\gamma^{\mu}\psi~+~eA_{\mu}\bar{\psi}\gamma^{\mu}\psi,
\end{equation} 
where $\psi$ is  fermionic field.
In the language of the Feynman graphs, we have in case of the vortical effects
two identical vertices and this suppresses the result by a factor of two compared
to the case of the chiral-magnetic effect when there are no identical vertices.
This is of course a common quantum effect.

 This factor of one half is absent from the counting the number of zero
 modes. The result for the vortical effect looks so as if there were no identical
 vertices. The reason is that the chemical potential $\mu$ plays two different roles.
 First, it determines the strength of the effective interaction, see (\ref{effective}).
 And second, the chemical potential determines the upper limit on
 the value of longitudinal component of the momentum of zero modes, see
 (\ref{upperlimit}). 
The factor $\mu^2$ which enters the final result (\ref{vortical})
is a combined effect of the strength of interaction, see (\ref{effective}),
and of the maximal value of the longitudinal momentum, see (\ref{upperlimit}). 
Thus, the source of the two factors $\mu$ is not identical now and, as a result,
 there is no
factor of one half.

To reiterate: the nowadays common derivation assumes that the liquid is characterized by a single velocity $u_{\mu}$ while 
the physical picture contains an additional component represented by the fermionic zero modes. 
Indeed, zero modes always propagate with speed of light and are not equilibrated
with the rest of the liquid. This appearance of the additional component does not reduce, to our mind, to the standard introduction of the normal component in the theory of superfluidity.   Remarkably, it does not affect
the final answer in case of the chiral magnetic effect because there are
no identical vertices in the triangle graph in this case.

  As mentioned above, in case of superfluidity the chiral current is
  indeed concentrated on the vortices. In this sense, the microscopic picture seems
  more reliable than the naive continuum limit (\ref{vortical}). In fact as the result of this note, for the case of superfluidity, the defect picture is the only one possible and the result (\ref{inmode}) is the one valid. 

In non-superfluid case the defect picture is also more reliable, generally speaking. However the theory is much less definite because of absence of vortex quantization.

  Thus, the dynamics of the vortices with zero fermionic modes living on them becomes
  the central issue to evaluate the chiral vortical effect.
  It is worth emphasizing, therefore,
that the model for the vortices considered here can well be oversimplified.
Indeed, in the picture considered all zero modes are peaked around the singular core
of a vortex. Thus, 4-fermionic interaction could be important. Moreover, in case of quarks
the effects of confinement can be crucial. 
 
\section{Aknowledgments}
We would like to acknowledge discussions with Z. Khaidukov and A.Krikun. The work of SAV and VPK was partly supported by the Dynasty foundation and FAIR program for master student. The work of VPK was also supported by DAAD Leonhard-Euler-Stipendium 2011-2012. The work of VIZ was partially supported by FEBR-11-02-01227-a and Fediral Special - Purpose Programme 'Cadres' of the Russian Ministry of Science and Education.

\section{Appendix}
In this section we explicitly construct the zero mode on the $n = 1$ vortex. As mentioned above, this problem reduces to the analysis of the 2-d Dirac operator given by:

 \begin{eqnarray}
 {\cal D^{\dag}} = -i \partial_1 + \partial_2   - (\partial_1 \varphi + i \partial_2 \varphi),
 \end{eqnarray}
here $\varphi$ stands for the polar angle, since the vortex field is given by $\phi = \mu t + \varphi$. If we substitute $\psi = \frac{f}{r}$  into 

 \begin{eqnarray}
 {\cal D^{\dag}} \psi = 0,
 \end{eqnarray}
we obtain for $f$
\begin{eqnarray}
(\partial_x + i \partial_ y) f = 0,
 \end{eqnarray}
since
\begin{eqnarray}
[-i (\partial_x + \frac{x}{r^2}) +  (\partial_y  + \frac{y}{r^2})] \frac{1}{r} = 0
\end{eqnarray}
 This  implies that $f$ is an entire function of $x + iy$. We now note that if $f$ is not a constant, then the norm of $\psi$ diverges at least power-like. It forces us to choose $f = 1$. The norm of $\psi$ still diverges, but only logarithmically. We can cut off this divergence on the upper limit by $r = R$, the radius of the bucket, and by $r = a$, the size of the vortex on the lower limit (it is noteworthy that nearby the vortex the equation $\phi = \mu t + \varphi$ is no longer valid anyway, so our consideration actually applies to distances $r >> a$). 

For  ${\cal D}$ one would analogically obtain by substituting $\psi = f r$ :
\begin{eqnarray}
(\partial_x - i \partial_ y) f = 0,
\end{eqnarray}
and $f$ is entire in $x - iy$. But for any choice of $f$ the norm of $\psi$ diverges. Thus we have obtained that $N_{+} = 1$, since there is one zero mode of $ {\cal D^{\dag}}$, and $N_{-} = 0$, and indeed $N = N_{+} - N_{-} = 1$. 
\\
\\
\\


\begin{thebibliography}{99}

\bibitem{erdmenger} J. Erdmenger, M. Haack, M. Kaminski,  A. Yarom,  JHEP 0901:055,2009.

\bibitem{sonsurowka} D.T. Son, P. Surowka,  Phys.Rev.Lett.103:191601,2009. 
 
\bibitem{kharzeev} K. Fukushima, D.E. Kharzeev, H.J. Warringa, Phys.Rev.D 78:074033,2008.


\bibitem{sonzhitnitsky} D.T. Son, A.R. Zhitnitsky, Phys.Rev.D 70:074018,2004.


\bibitem{zahed} M. Lublinsky, I. Zahed, Phys.Lett.B 684:119-122,2010.

\bibitem{sadofyev} M.I. Isachenkov, A.V. Sadofyev, Phys.Lett.B 697:404-406,2011.

\bibitem{zakharov} A.V. Sadofyev, V.I. Shevchenko, V.I. Zakharov, Phys.Rev.D83:105025,2011.

\bibitem{Tigran} I. Gahramanov, T. Kalaydzhyan, I. Kirsch, arXiv:1203.4259 [hep-th].

\bibitem{Roy} V.P. Nair, R. Ray, S. Roy, arXiv:1112.4022 [hep-th].

\bibitem{new} N. Banerjee, J. Bhattacharya, S. Bhattacharyya, S. Jain, S. Minwalla, T. Sharma, arXiv:1203.3544 [hep-th];  K. Jensen, arXiv:1203.3599 [hep-th].

\bibitem{goldstonewilczec} J. Goldstone, F. Wilczek,  Phys.Rev.Lett. 47:986-989,1981.

\bibitem{nielsen} H. B. Nielsen and M. Ninomiya, Phys.Lett. 130B,389,1983.

\bibitem{callanharvey} C.G. Callan, J.A. Harvey, Nucl.Phys.B 250:427,1985.
 
\bibitem{zhitnitsky} M.A. Metlitski, A.R. Zhitnitsky, Phys.Rev.D 72:045011,2005.

\bibitem{landau6}  L. D. Landau and E. M. Lifshitz, Fluid Mechanics, Pergamon, New York (1959).

\bibitem{sonstephanov} D.T. Son, M.A. Stephanov, Phys.Atom.Nucl.64:834-842,2001, Yad.Fiz.64:899-907,2001;\\ J.B. Kogut, M. A. Stephanov, D. Toublan, Phys.Lett.B 464:183-191,1999.

\bibitem{zamaklar} O. Aharony, K. Peeters, J. Sonnenschein, M. Zamaklar, JHEP 0802:071,2008. 

\bibitem{sonsup} D. T. Son, arXiv:0204199v2 [hep-ph].

\bibitem{Landau9} L. D. Landau and E. M. Lifshitz, Statistical Physics, Part 2, Pergamon, New York (1959).

\bibitem{nicolis} A. Nicolis arXiv:1108.2513v1 [hep-th]

\bibitem{Casher} Y. Aharonov, A. Casher, Phys.Rev.A 19:2461-2462,1979.

\bibitem{semenoff} A.J. Niemi, G.W. Semenoff, Phys.Rept.135:99,1986.




\end{thebibliography}
\end{document}